\documentclass[conference]{IEEEtran}
\IEEEoverridecommandlockouts
\usepackage{cite}
\usepackage{amsmath,amssymb,amsfonts}
\usepackage{algorithmic}
\usepackage{algorithm}
\usepackage{graphicx}
\usepackage{textcomp}
\usepackage{xcolor}
\usepackage{hyperref}
\usepackage{url}
\usepackage[subrefformat=parens]{subcaption}
\usepackage{siunitx}
\newcommand{\figref}[1]{Fig.\,\ref{#1}}
\newcommand{\Figref}[1]{Figure\,\ref{#1}}
\newcommand{\tabref}[1]{Table\,\ref{#1}}
\newcommand{\eref}[1]{Eq.\,(\ref{#1})}
\newcommand{\secref}[1]{Sec.\,\ref{#1}}
\newcommand{\ie}{\emph{i.e.}}

\newcommand{\rev}[1]{#1} 

\newcommand{\mE}{\mathbb{E}}

\begin{document}

\title{\rev{Sizing of Battery Considering Renewable Energy Bidding Strategy with Reinforcement Learning}
\thanks{This work was supported in part by JST, ACT-X Grant Number JPMJAX210M, Japan, and in part by the Kansai Research Foundation for Technology Promotion, Japan.}
}

\author{
\IEEEauthorblockN{Taiyo Mantani, Hikaru Hoshino, Tomonari Kanazawa, Eiko Furutani}
\IEEEauthorblockA{\textit{Department of Electrical Materials and Engineering} \\
\textit{University of Hyogo}\\
Himeji, Japan \\
\{er24y025@guh, hoshino@eng, eo21u029@guh, furutani@eng\}.u-hyogo.ac.jp}
}

\maketitle

\begin{abstract}
This paper proposes a novel computationally efficient algorithm for \rev{optimal sizing of Battery Energy Storage Systems (BESS) considering renewable energy bidding strategies.
Unlike existing two-stage methods, our algorithm enables the co-optimization of both} by updating the BESS size during the training of the bidding policy, leveraging an extended reinforcement learning (RL) framework inspired by advancements in embodied cognition. By integrating the Deep Recurrent Q-Network (DRQN) with a distributed RL framework, the proposed algorithm effectively manages uncertainties in renewable generation and market prices while enabling parallel computation for efficiently handling long-term data.


\end{abstract}


\section{Introduction}

With the growing penetration of Variable Renewable Energy (VRE) sources such as photovoltaic (PV) and wind power, the market integration of VRE~\cite{Hu2018} is becoming an important factor to minimize market distortions and drive the development of new business models~\cite{Huntington2017,Botelho2021}. 
Renewable producers are expected to participate in electricity markets and respond to short-term price signals as much as possible. 
Battery Energy Storage Systems (BESSs) provide operational flexibility to avoid the penalty of not satisfying the dispatch commitment and reducing generation curtailment~\cite{Lin2024}. 
Furthermore, energy arbitrage by charging energy when the price is low and discharging energy when the price is high can result in additional net revenue for renewable producers~\cite{Huang2021}.  

Many studies investigated market bidding and BESS control strategies to enhance the total profits of renewable producers. 
One of the main challenges in this kind of operational optimization problem is modeling the uncertainty of renewable generation and energy prices. 
In this regard, deep Reinforcement Learning (RL) has been widely used (see e.g. \cite{Ye2020,Cao2020,Jeong23:DRL_energy}) and shows better performance than mathematical optimization techniques such as stochastic programming or robust optimization, since the former requires simulation of numerous scenarios resulting in a significant computational burden~\cite{Jeong23:DRL_energy}, and the latter may be too conservative since it is intrinsically designed to be sub-optimal~\cite{Jeong23:DRL_energy}.

Meanwhile, optimal sizing of BESS has been studied in different scenarios as reviewed in \cite{Zhou2024:Storage_sizing}. 
However, existing methods do not necessarily consider market bidding or BESS control strategies effectively. 
On one hand, most of the existing methods use oversimplified models based on rules of thumb or do not precisely account for important factors such as market dynamics~\cite{Lin2024}.
On the other hand, \rev{sizing of BESS considering optimal bidding strategy} usually requires prohibitively heavy computational loads to obtain global optimization results~\cite{Zhou2024:Storage_sizing}. 
To overcome this difficulty, recent studies have begun to explore machine learning-based approaches~\cite{Lin2024,Kang2023,Li2022}. 
In these studies, the role of machine learning is to construct a surrogate model for accurate prediction of operational performance of energy systems in supervised~\cite{Lin2024} or RL~\cite{Kang2023,Li2022} settings. 
The trained models are then integrated with 
a standard optimization algorithm for storage sizing. 
However, with this kind of two-stage approaches, the trained model must be accurate over the entire search space, and the training process may require huge computational resources to prepare a complete training data set. 


This paper presents a computationally efficient method for \rev{sizing of BESS considering renewable energy bidding strategies. 
In contrast to existing two-stage methods \cite{Lin2024,Kang2023,Li2022}, we propose a co-optimization} algorithm where the BESS size is optimized during the training of bidding strategy, by extending a standard RL algorithm for the purpose of co-optimization.
Although similar algorithms have been proposed in the field of embodied cognition~\cite{Schaff2019,Chen2021}, where the physical design of a robot and the policy to control its motion are simultaneously optimized, our algorithm is novel in that it is based on the Deep Recurrent Q-Network (DRQN)~\cite{Hausknecht2015:DRQN} combined with a distributed RL framework as proposed in Ape-X~\cite{Horgan2018:Ape-X} \rev{to manage uncertainties in renewable generation and market prices in short-term operations and to efficiently deal with long-term generation/market data for optimal BESS sizing}. 
To the best of our knowledge, this is the first work to use this kind of co-optimization algorithm for energy systems application. 
Our implementation is available at \url{https://github.com/uhyogo-epa/IEEE_PESGM_2025_mantani_RE}. 

The rest of this paper is organized as follows. 
In \secref{sec:market_model}, we introduce a renewable energy bidding model for a general real-time market. In \secref{sec:algorithm}, we describe the proposed co-optimization method. 
In \secref{sec:simulation}, numerical experiments are given followed by conclusion in \secref{sec:conclusion}.

\section{Renewable Energy Bidding Model} \label{sec:market_model}
    
\begin{figure}[t!]
\centering
\includegraphics[width=\linewidth]{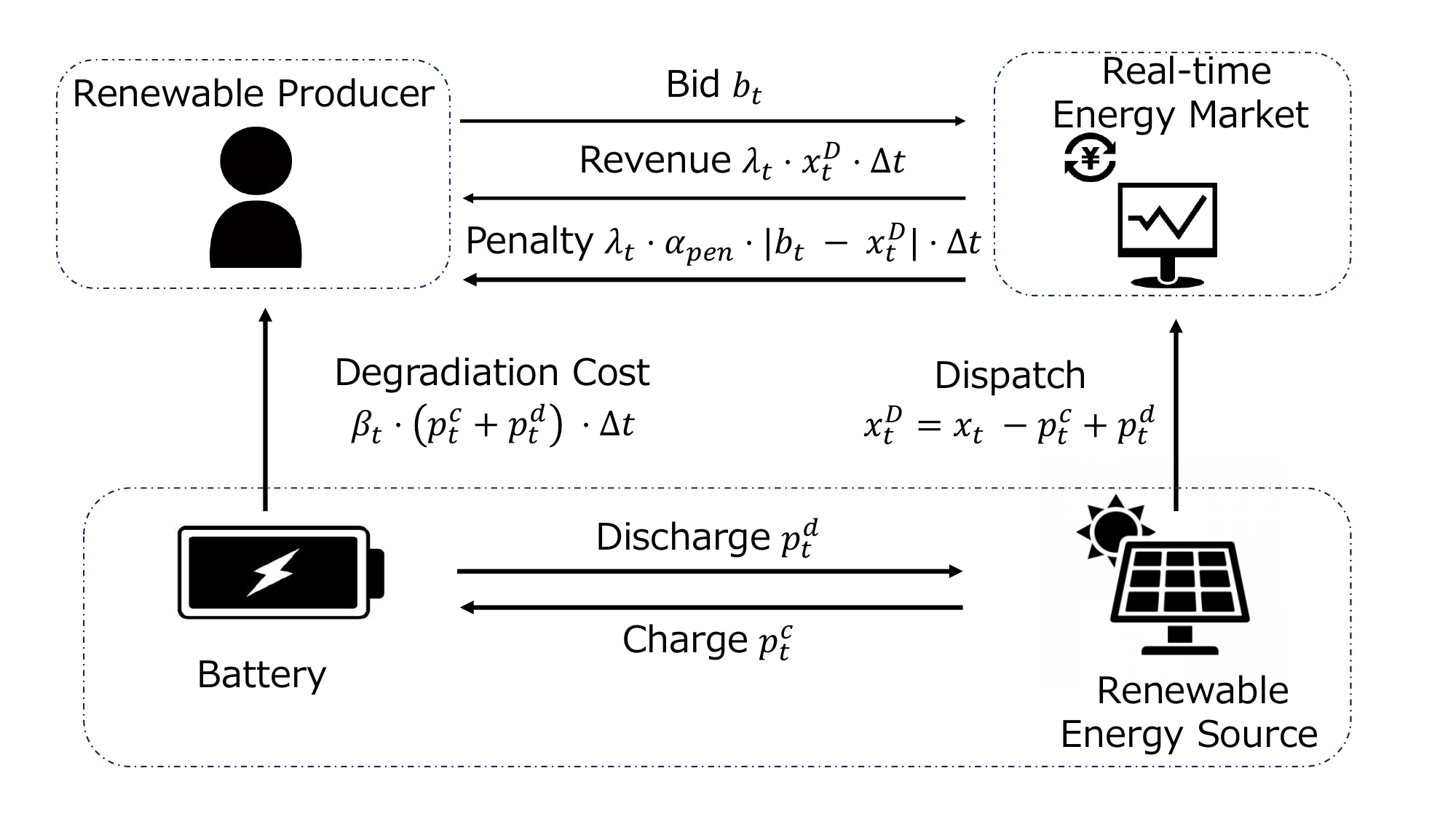}
\caption{Structure of real-time energy market process}
\label{fig:market_structure}
\vspace{-6mm}
\end{figure}

We first review a renewable energy bidding model based on \cite{Jeong23:DRL_energy}. 
This model represents a generic structure of real-time markets where a renewable producer tries to maximize its total profit through energy trading as shown in \figref{fig:market_structure}. 
The dispatched power $x_t^\mathrm{D}$ at time $t$ can be  described as 
\begin{equation}
   x^\mathrm{D}_t = x_t - p^\mathrm{c}_t + p^\mathrm{d}_t,
   \label{eq:supply_energy}
\end{equation}
where $x_t$ stands for the generation output of renewable sources, and $p_t^\mathrm{c} (\geq 0)$ and $p_t^\mathrm{d} (\geq 0)$ for the charged power and discharged power of the battery, respectively. 
Here, all units are in megawatts (\si{MW}). 
The market gate closure is one time step prior, meaning that at time $t-1$, the renewable producer bids $b_t$ based on predictions of $x_t$ and the market price $\lambda_t$ for the next time $t$. 
Any discrepancy between $b_t$ and $x_t^\mathrm{D}$ is subject to imbalance payments, and by letting $\lambda^\mathrm{sur}_t (< \lambda_t)$ and $\lambda^\mathrm{def}_t (> \lambda_t)$ represent the surplus and deficit imbalance settlement prices, respectively, the revenue $F_t^\mathrm{m}$ of 
the renewable producer at time $t$ is given as follows:
\begin{align}
   F^\mathrm{m}_t = & \lambda_t b_t \Delta t  
   +  \left( \lambda^\mathrm{sur}_t (x^\mathrm{D}_t - b_t)^+ - \lambda^\mathrm{def}_t (b_t-x^\mathrm{D}_t)^+ \right) \Delta t 
   \label{eq:revenue_original}
\end{align}
where $(z)^+ := \max(z,0)$ for any $z$, and $\Delta t$ stands for the duration of time slot. 
If we assume $\lambda^\mathrm{sur}_t = (1-\alpha_\mathrm{pen})\lambda_t$ and $\lambda^\mathrm{def}_t = (1+\alpha_\mathrm{pen})\lambda_t$ with a penalty factor $\alpha_\mathrm{pen}$,  \eref{eq:revenue_original} can be rewritten as 
\begin{align}
  F^\mathrm{m}_t = \lambda_t ( x^\mathrm{D}_t - \alpha_\mathrm{pen} |b_t - x^\mathrm{D}_t| ) \Delta t. \label{eq:Fm}
\end{align}

To maximize the renewable producer's profit, the bid $b_t$ and the BESS control need to be optimized while satisfying operational constraints. 
The State of Charge (SOC) of the BESS, denoted by $x^\mathrm{soc}_t$, should be constrained in the upper and lower limits, $S^\mathrm{max}$ and  $S^\mathrm{min}$, respectively: 
\begin{align}
 0 < S^\mathrm{min} \le x^\mathrm{soc}_t \le S^\mathrm{max} < 1, 
 \end{align}
and the time evolution of the SOC is given by
\begin{equation}
    x^\mathrm{soc}_{t+1} = x^\mathrm{soc}_t + \eta^\mathrm{c}_t\frac{p^\mathrm{c}_t}{E_\mathrm{max}}\Delta{t} - \frac{1}{\eta^\mathrm{d}_t}\frac{p^\mathrm{d}_t}{E_\mathrm{max}} \Delta t,\label{eq:battery_model}
\end{equation}
where $\eta^\mathrm{c}_t$ and $\eta^\mathrm{d}_t$ stands for the charging and discharging efficiencies, respectively, and $E_\mathrm{max}$ for the BESS size (battery capacity).
The maximum charging and discharging limits are assumed to be proportional to the BESS size $E_\mathrm{max}$: 
\begin{align}
    p_t^\mathrm{c} \le P^\mathrm{c,unit}_\mathrm{max} E_\mathrm{max}, \quad 
    p_t^\mathrm{d} \le P^\mathrm{d,unit}_\mathrm{max} E_\mathrm{max}
\end{align}
with the per unit upper limits $P^\mathrm{c,unit}_\mathrm{max}$ and $P^\mathrm{d,unit}_\mathrm{max}$. 
In addition to these, we consider a battery degradation cost $\beta_t$, and the objective is to maximize the total net profit $F$ given by
\begin{align}
  \label{eq:total_revenue}
   F = \sum_{t \in \mathcal{T}} F_t = \sum_{t \in \mathcal{T}} \underbrace{ \left( F^\mathrm{m}_t -
    \beta_t(p^\mathrm{c}_t + p^\mathrm{d}_t) \Delta{t} \right) }_{:= F_t}
\end{align}
where $F_t$ represents the net profit obtained at time $t$, and $\mathcal{T}$ represents the set of time indices under consideration.

\section{Co-optimization Algorithm} \label{sec:algorithm}

This section presents the proposed algorithm.  
An RL-based bidding strategy optimization framework is introduced in \secref{sec:rl_bidding}. 
We propose an extension of the framework for the co-optimization in \secref{sec:co-optimization_alg}. 

\subsection{Bidding Strategy by Reinforcement Learning} \label{sec:rl_bidding}

The objective of the training is to maximize the sum of the profit $F=\sum F_t$ over all time slots.
Although traditional mathematical optimization techniques such as stochastic programming could be used, many previous studies find that Deep RL shows better performance under energy price and/or renewable generation uncertainty (see~\cite{Ye2020,Cao2020,Jeong23:DRL_energy}, and references therein). 
In the context of RL, an agent interacts with an environment by sequentially taking actions and observing instantaneous rewards and state transitions over a sequence of time steps to maximize a cumulative reward with a discount $\gamma$.
Since the renewable generation $x_t$ and the real-time energy market price $\lambda_t$ are not known in advance, we consider the observation at time $t$ to be $o_t := (x_{t-1}, \lambda_{t-1}, x^\mathrm{soc}_t)$ and construct the state by
\begin{equation}
    s_t = (o_0, o_1, ... , o_{t-1}, o_t). \label{eq:state}
\end{equation}
Then, the problem of determining optimal action $a_t$ from the observation $o_t$ can be treated as a decision-making problem for a Partially Observable  Markov Decision Process (POMDP). 
To solve this, the Deep Recurrent Q-Network (DRQN)~\cite{Hausknecht2015:DRQN} is available and widely used in the field of RL research. 
DRQN integrates deep Q-learning with a Long Short-term Memory (LSTM) and estimates the action-value function in the form of $Q(o_t, h_t, c_t, a_t)$, where $h_t$ and $c_t$ stand for the hidden and cell states of the LSTM, respectively. 

The agent’s decision variable consists of the bidding value $b_t$, the charging power $p^{\mathrm{c}}_t$, and the discharging power $p^{\mathrm{d}}_t$.  
For the purpose of error compensation, excessive power $x_t-b_t$ can be charged in the battery when $b_t < x_t$, and the power deficit $b_t-x_t$ can be discharged from the battery when $b_t > x_t$. 
Furthermore, for energy arbitrage, charging or discharging power can be increased (or decreased) by the scaling factor $\rho_t$ considering the price $\lambda_t$ \cite{Jeong23:DRL_energy}: charging more power when $\lambda_t$ is low (or negative) and discharging more power when $\lambda_t$ is high.
By following~\cite{Jeong23:DRL_energy}, we consider the action of the agent as 
\begin{align}
    a_t = ( b_t, \, \rho_t ),
\end{align}
and the corresponding $p^\mathrm{c}_t$ and $p^\mathrm{d}_t$ are determined as
\begin{subequations}
\begin{align}
    p^\mathrm{c}_t = \min((\rho_t(x_t-b_t))^+, \bar{P}^\mathrm{c}_t), \\
    p^\mathrm{d}_t = \min((\rho_t(b_t-x_t))^+, \bar{P}^\mathrm{d}_t), 
\end{align}
\end{subequations}
where $\bar{P}^\mathrm{c}_t$ and $\bar{P}^\mathrm{d}_t$ represent the upper limits of charging and discharging power considering the upper and lower limits of the SOC, $x^\mathrm{soc}_t$, respectively, and are given by
\begin{align}
    \bar{P}^\mathrm{c}_t := \min\left( \dfrac{E_\mathrm{max}(S^\mathrm{max} - x^\mathrm{soc}_t)}{\Delta t} , \, P^\mathrm{c}_\mathrm{max} \right), \\
    \bar{P}^\mathrm{d}_t := \min\left( \dfrac{E_\mathrm{max}( x^\mathrm{soc}_t - S^\mathrm{min})}{\Delta t} , \, P^\mathrm{d}_\mathrm{max} \right). 
\end{align}
For the purpose of variance reduction in learning process~\cite{Jeong23:DRL_energy}, the reward $r_t$ is formulated as 
\begin{equation}
    r_t = F_t - \lambda_t {x_t} \Delta{t}. \label{eq:reward}
\end{equation}
Here, $x_t$ and $\lambda_t$ are determined independently of $a
_t$, and $r_t$ increases monotonically with respect to $F_t$, which implies that maximizing $r_t$ is equivalent with the maximizing $F_t$.

\subsection{Co-optimization of Energy Bidding and Battery Size} \label{sec:co-optimization_alg}

Standard RL trains a policy for an agent with a fixed system design, but this design affects the optimal policy and the long-term reward. 
In the context of renewable energy bidding, the battery size affects the bidding strategy and the profit of the renewable producer. 
We propose an extension of the aforementioned RL framework to co-optimize the battery size as well as the bidding strategy. 
For this, we denote the system's design parameter by $\omega$, \ie, $\omega=  \{ E_\mathrm{max} \}$, and consider an augmented Q-function denoted by $Q(o_t, h_t, c_t, a_t, \omega)$ with its neural-network approximator $\hat{Q}_\theta(o_t, h_t, c_t, a_t, \omega)$ with weights $\theta$. 
Then, the design $\omega$ can be considered as an additional action fixed over an entire episode and generated by a (state independent) policy with a distribution $p_\mu(\omega)$. 
Here, the parameter $\mu$ is learnable and encodes the belief about which designs are likely to be successful. 
In this paper, we use a Gaussian distribution where its mean is $\mu$ (learnable), and the variance $\sigma^2$ is fixed. 

The proposed algorithm for updating $\theta$ and $\mu$ is summarized in Algorithm\,1 as a high-level pseudo code. 
The overall structure follows from the standard RL framework, and the main loop is iterated over episodes. 
At the beginning of each episode, the design $\omega$ is sampled from the distribution $p_\mu$ (line 3), and the parameter $\theta$ is updated by using data stored in the experience memory \cite{Hausknecht2015:DRQN} (lines 4 and 5). 
While DRQN is used in this paper, other RL algorithms can also be used for updating the parameter $\theta$. 
Also, data collection in lines 3 and 4 can be parallelized using multiple workers as in Ape-X \cite{Horgan2018:Ape-X}. 
Then, the parameter $\mu$ is updated once in every $N_\mathrm{up}$ episodes. 
Following the policy gradient theorem \cite{Sutton2018:RL}, the gradient of the expected return  with respect to $\mu$ can be given by 
\begin{align} \label{eq:gradient_mu}
 \nabla_\mu \mE[G] = \mE \left[ \nabla_\mu \ln p_\mu(\omega) G
\right]
\end{align}
with the return $G$ given by
\begin{align}
    \label{eq:net_profit_G}
    G = W_\mathrm{anu} \sum_t r_t - \omega P_\mathrm{bat}
\end{align}
where $W_\mathrm{anu}$ stands for the annualization factor, and $P_\mathrm{bat}$ for the BESS installation cost per unit capacity per year. 
This can be approximated by the following empirical estimator using the design $\omega_i$ and the return $G_i$ for the episode $i$:
\begin{align}
    \nabla_\mu \mE[G] \approx \sum_{i \in \mathcal{E}} \dfrac{\omega_i - \mu}{\sigma^2} ({G}_i - \bar{G})
    \label{eq:gradient_G}
\end{align}
where $\mathcal{E}$ stands for the indices of previous $N_\mathrm{up}$ episodes and $\bar{G}$ for the mean of $G_i$ over $\mathcal{E}$.

\begin{figure}[!t]
  \begin{algorithm}[H]
  \vspace{-3mm}
      \caption{Co-optimization of bidding and battery size}
      \label{alg:UCCD}
      \begin{algorithmic}[1]
      \STATE Initialize $\widehat{Q}_\theta$ with weights $\theta$ and $p_\mu$ with mean $\mu$
      \FOR{episode = 1:$N$}
      \STATE Sample design $\omega \sim p_\mu$
      \STATE Observe trajectory $\{ o_1, a_1, r_1, \dots, a_{T-1}, r_{T-1}, o_T \}$ for $T$ time steps and store in experience memory
      \STATE  Update $\theta$ using DRQN and experience memory
      \IF {episode\,\%\,$N_\mathrm{up}$ == 0 }
      \STATE Compute return $G$ in previous $N_\mathrm{up}$ episodes
      \STATE Update $\mu$ by using \eqref{eq:gradient_G}
      \ENDIF
      \ENDFOR
      \end{algorithmic}
  \end{algorithm}
  \vspace{-5mm}
\end{figure}


\section{Numerical Experiments}
\label{sec:simulation}

This section presents numerical results of co-optimization. 
The effectiveness of the proposed method is verified regarding its accuracy and computational time by comparison with a standard two-stage framework in \secref{sec:comparison}. 
This is done by using short-term data, and the applicability of the proposed method to the long-term data is examined in \secref{sec:long_term}.  


\subsection{Verification Using Short-term Data}
\label{sec:comparison}

\begin{figure}[t!]
\centering
\begin{minipage}[b]{1.0\columnwidth}
    \centering
  \includegraphics[width=0.95\columnwidth]{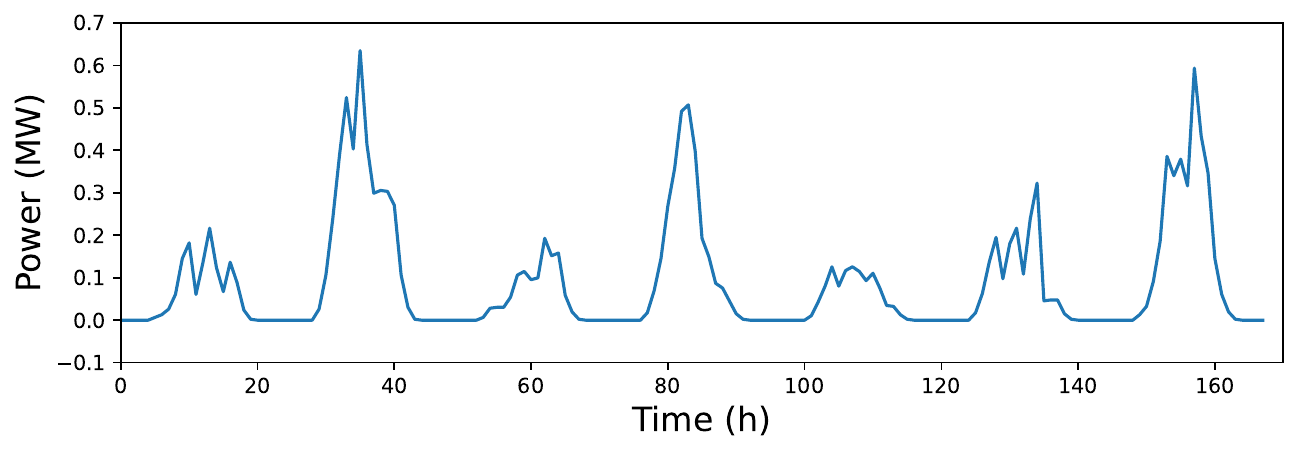}
  \vspace{-2mm}
    \subcaption{Solar generation}
    \label{fig:solar_generation}
\end{minipage}
\begin{minipage}[b]{1.0\columnwidth}
    \centering
\includegraphics[width=0.95\columnwidth]{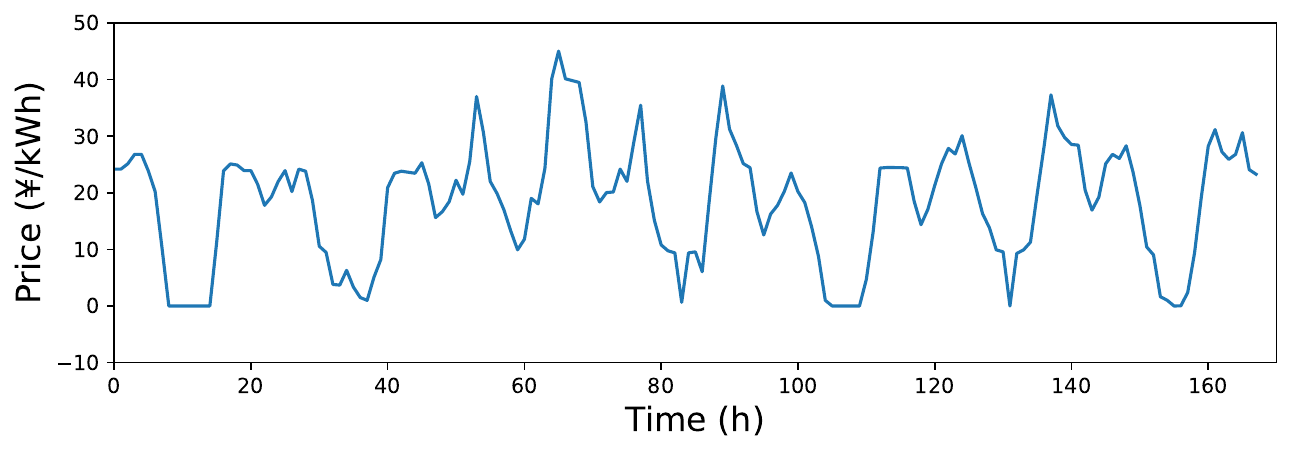}
    \subcaption{Market Price}
    \label{fig:market_price}
   \end{minipage}
\caption{Time series data used for simulation} \label{fig:time_series}
\vspace{-6mm}
\end{figure}

In this subsection, the proposed method is verified with data length of one week as shown in \figref{fig:time_series}. 
The PV generation profile is based on solar irradiance data starting from July 1 of an average year obtained from METPV-20 \cite{METPV}. 
Market prices are taken from one week of system price data in the JEPX \cite{JEPX} spot market, beginning July 1, 2022.
The parameter settings for the numerical simulations are shown in \tabref{tab:parameter}. 
For DRQN, a neural network consisting of three layers of a fully connected, an LSTM, and a fully connected layers (each with 64 units) was used. 
The update method employed Bootstrapped Random Updates, and the model was trained with Adam with the learning rate of $\SI{1e-4}{}$ for the DRQN weights and $\SI{1e-5}{}$ for $\mu$.
The parameter $\sigma$ was fixed to $0.2$.
The bid $b_t$ was from $\SI{0}{MW}$ to $\SI{0.6}{MW}$ in increments of $\SI{0.1}{MW}$, while the scaling factor $\rho_t$ was one of $0.0$, $0.5$, and $1.0$. 
The penalty factor was $\alpha_\mathrm{pen}=1.0$. 
The discount rate during training was set to 0.9.
\rev{The calculations were performed using a computer with an Intel Core i7-9700K CPU.}

\begin{figure}[t!]
 \centering
 \includegraphics[width=\linewidth]{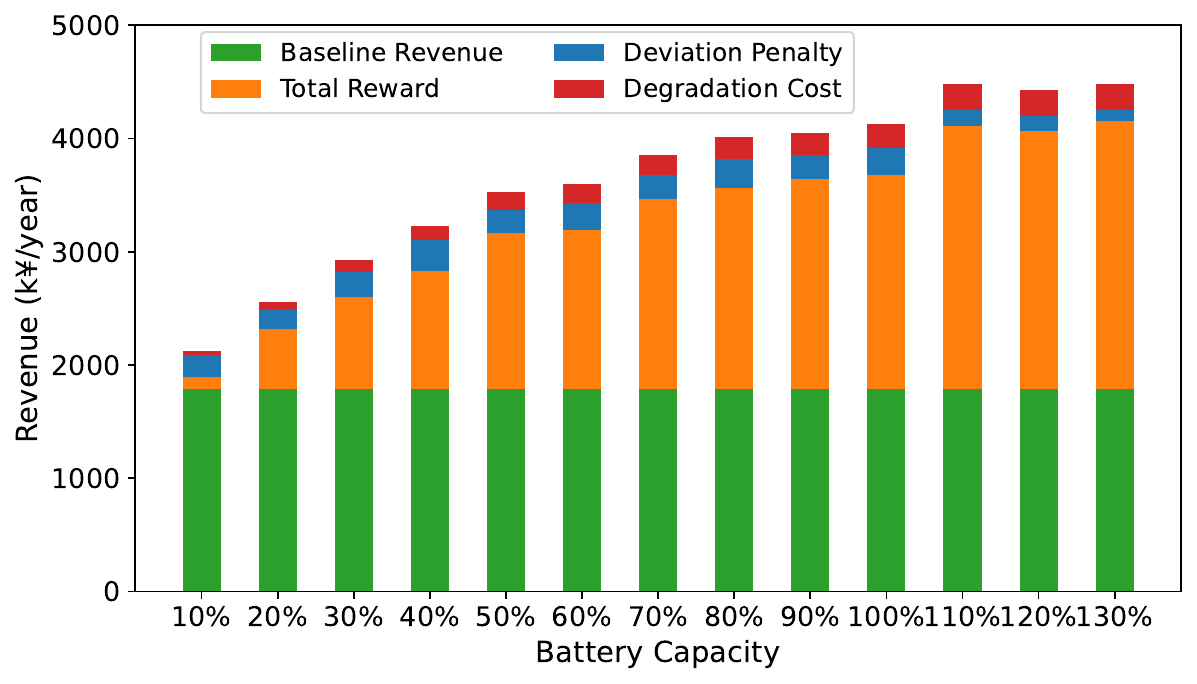}
 \caption{Total net revenue $F$ for various BESS sizes}
 \label{fig:bar_graph}
\end{figure}

\begin{figure}[t!]
 \centering
 \vspace{-3mm}
 \includegraphics[width=\linewidth]{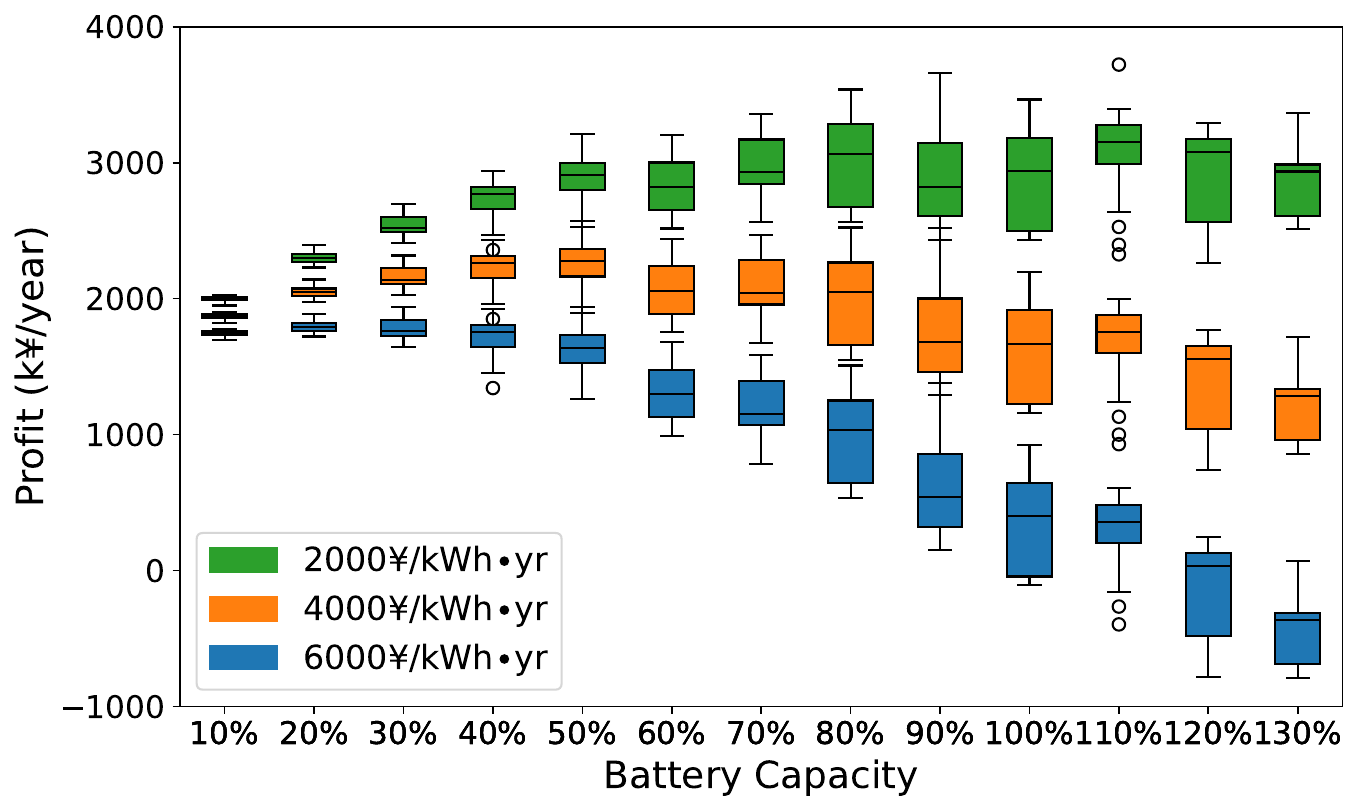}
 \caption{Net profit $G$ considering BESS installation costs}
 \label{fig:box_graph_2}
 \vspace{-4mm}
\end{figure}

For comparison, we first analyze the optimal BESS size in a standard two-stage framework. 
The BESS size $E_\mathrm{max}$ was varied from $\SI{10}{\%}$ to $\SI{130}{\%}$ of the hourly electricity generation with the maximum solar irradiance, and a standard DRQN agent was trained for each setting of $E_\mathrm{max}$. 
\Figref{fig:bar_graph} shows the (annualized) total revenue that the renewable producer obtains from the market trading after 800 episodes of training \rev{(around 8 minutes in average)}. 
Since RL training sometime failed, we performed 18 repeated experiments and excluded 3 cases that did not perform well to plot averaged results for the remaining 15 runs. 
In the figure, \emph{Baseline Revenue} represents the revenue when the generated electricity is directly sold to the market, \ie, $\sum \lambda_t x_t\Delta t$, and \emph{Total Reward} represents $\sum r_t$ with the reward $r_t$ given in \eqref{eq:reward}. 
\rev{The sum of these two corresponds to the net revenue $F= \sum r_t + \lambda_t x_t\Delta t$ defined in \eqref{eq:total_revenue}. 
Also, \emph{Deviation Penalty} represents the penalty term in  \eqref{eq:Fm}, \ie, $\sum_t \alpha_\mathrm{pen} |b_t - x^\mathrm{D}_t|$, and \emph{Degradation Cost} shows the total battery degradation cost in \eqref{eq:total_revenue}, \ie, $\sum_t \beta_t(p^\mathrm{c}_t + p^\mathrm{d}_t) \Delta{t} $.}
As an overall trend, it can be seen that the net revenue $F$ increases as the BESS size increases. 
Furthermore, \figref{fig:box_graph_2} shows the net profit $G$ defined as \eqref{eq:net_profit_G} in a box plot for each setting of the BESS price of $\SI{2000}{\yen/(kWh. yr)}$, $\SI{4000}{\yen/(kWh.yr)}$, and $\SI{6000}{(\yen/kWh.yr)}$, and the optimal BESS size can be determined as 110--120\,$\si{\%}$, 40--50\,$\si{\%}$, and 20--30$\si{\%}$, respectively. 

\begin{table}[t!]
\centering
\renewcommand{\arraystretch}{1.05}
\caption{Parameters for numerical simulation}
\label{tab:parameter}
\begin{tabular}{|c|c|c|}
\hline
 Symbol & Parameter & Values \\
 \hline
$\mathrm{SOC}_{\min}$ & minimum value of SOC  & 0.1 \\
$\mathrm{SOC}_{\max}$ & maximum value of SOC  & 0.9 \\
$\eta_t^\mathrm{c}$ &  charge efficiency  & 0.96 \\
$\eta_t^\mathrm{d}$ &  discharge efficiency  & 0.995\\
$\alpha_\mathrm{pen}$ &  penalty for imbalance  &  1.0 / 0.5\\
$\beta_t$ &  Degradation cost of battery  & ¥1.0/MWh \\
 \hline
 \end{tabular}
\end{table}

\begin{figure}[t!]
 \centering
 \includegraphics[width=\linewidth]{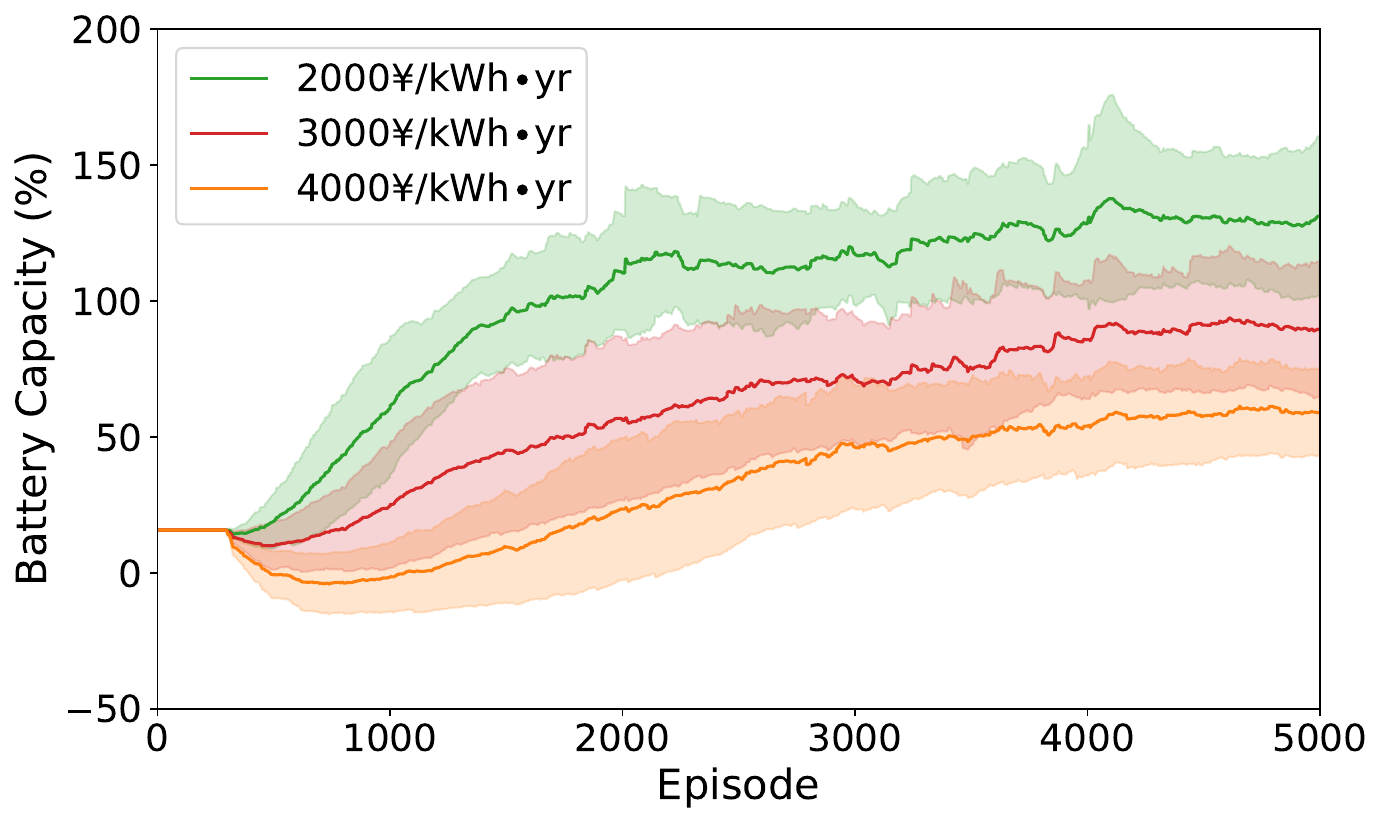}
 \caption{Change in the value $\mu$ during learning process}
 \label{fig:update_mu} 
  \vspace{-2mm}
\end{figure}

Next, the results of the proposed co-optimization algorithm is shown in \figref{fig:update_mu}. 
The solid lines in the figure show the average of the value $\mu$ of 15 repeated experiments and the shaded regions show their standard deviations. 
With the setting of $\SI{2000}{\yen/(kWh. yr)}$ (shown by \emph{green}) and $\SI{4000}{\yen/(kWh.yr)}$ (shown by \emph{orange}), it can be confirmed that the value of $\mu$ converges to around $\si{120}{\%}$ and $\si{50}{\%}$, respectively, and is consistent with the results in \figref{fig:box_graph_2}.
While not presented in \figref{fig:box_graph_2}, the optimal BESS size for the setting of $\SI{3000}{\yen/(kWh. yr)}$ (shown by \emph{red}) has been identified as around 70--80\,$\si{\%}$, and the co-optimization result is consistent for this setting. 
For $\SI{6000}{\yen/(kWh.yr)}$, the value $\mu$ kept decreasing during training process, and the optimal value was not obtained with the current setting.
Nevertheless, the proposed algorithm worked in optimizing the BESS size ranging from $\SI{50}{\%}$ to $\SI{120}{\%}$. 
Regarding the computational time, it took about 4,000 episodes until convergence \rev{(around 48 minutes in average)} 
This is 6 times longer than when $\mu$ is fixed, where it took 600 to 800 episodes for convergence, but it is faster than determining the BESS size by exhaustively calculating the net profits as in \figref{fig:box_graph_2}. 

\subsection{Applicability to Long-Term Data}
\label{sec:long_term}

\begin{figure}[t!]
 \centering
 \vspace{-1mm}
 \includegraphics[width=\linewidth]{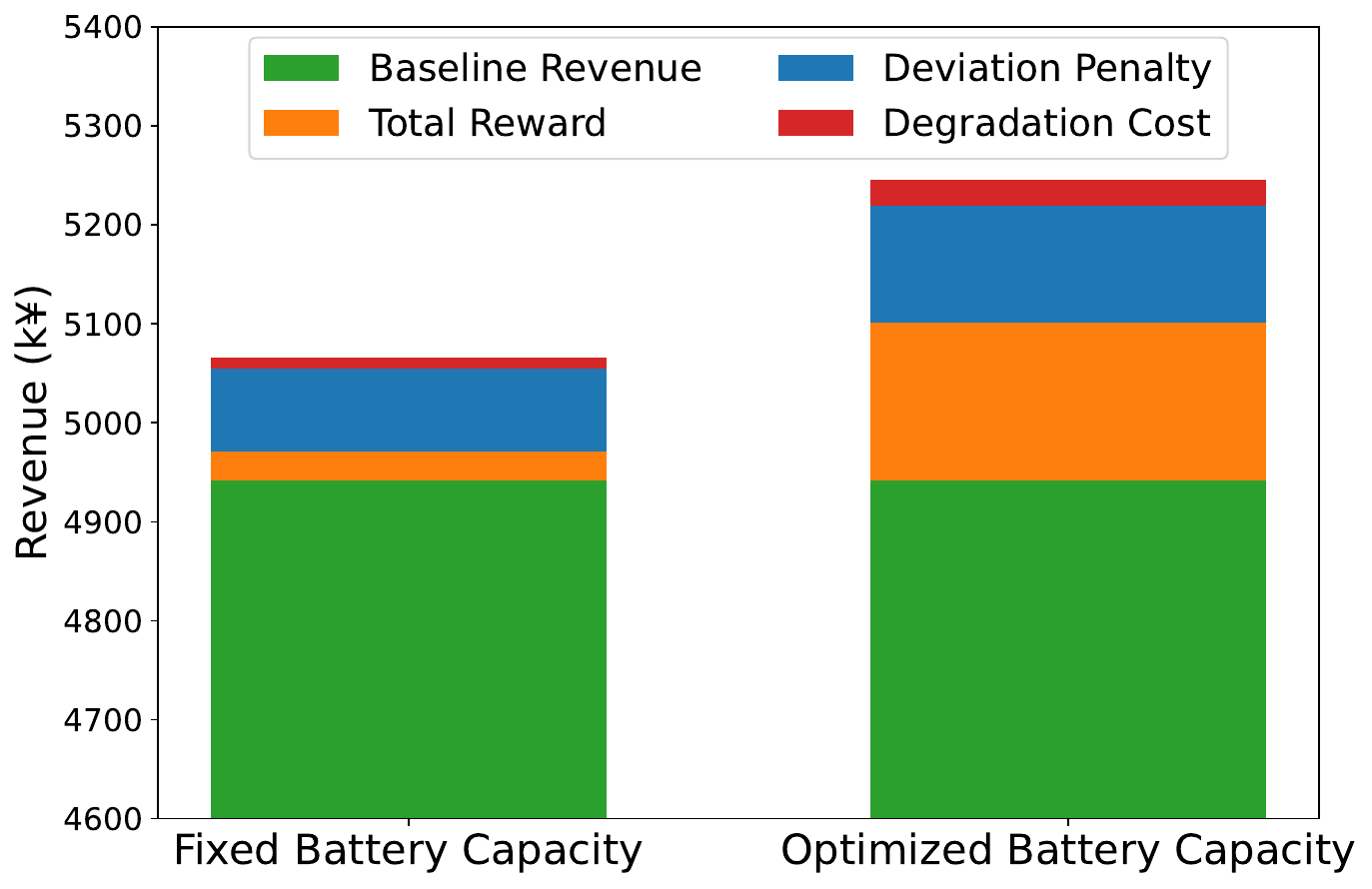} 
 \caption{Total revenues with fixed and optimized BESS sizes}
 \label{fig:bar_graph_year}
  \vspace{-3mm}
\end{figure}

Here, we present the results using data for the entire period in 2022. 
For this, we parallelized workers as in Ape-X~\cite{Horgan2018:Ape-X} and used 12 workers to obtain one month of training data for each worker. 
The bid $b_t$ ranged from $\SI{0}{MW}$ to $\SI{0.8}{MW}$ to cover the maximum power generation, and the scaling factor $\rho_t$ was one of $0.0$, $0.5$, $1.0$, and $1.5$. The penalty factor was set to $\alpha_\mathrm{pen}=0.5$. 
The BESS price was set at $\SI{1000}{\yen/(kWh. yr)}$. 
Compared to training using one-week data in \secref{sec:comparison}, training was more difficult when using long-term one-year data. 
This can be partly due to variations in seasonal trends in the studied data and partly due to the difficulty in learning policy that is effective for long episodes. 
To cope with the latter, the learning rate for the DRQN weights was changed to $\SI{1e-3}{}$, and the parameter $\sigma$ was changed to $0.15$.
The discount rate was set to 0.99. 
\rev{Each training process took approximately 11 hours on a workstation with an AMD EPYC 7763 CPU and an NVIDIA RTX 6000Ada GPU.}
\Figref{fig:bar_graph_year} shows the total revenues after 20,000 episodes when the BESS size is fixed to the initial setting (\emph{Fixed Battery Capacity}) and when the BESS size is learned using the proposed co-optimization method (\emph{Optimized Battery Capacity}). 
In the figure, the maximum revenue obtained from 8 repeated experiments is shown for each of the above two cases. Among these experiments, the
total reward became negative (training failed) in 7 out of 8
experiments (87.5\%) when the BESS size was fixed, but was
negative in 3 experiments (37.5\%) when the co-optimization
method is used. Thus, co-optimizing the RL policy and the
BESS size is also beneficial for avoiding failure in the training
of policy by exploring better system configuration (the BESS
size) simultaneously.

\section{Conclusions}
\label{sec:conclusion}

This paper proposed a co-optimization method for RL-based bidding policy and BESS sizing. 
The proposed method is computationally efficient compared with exhaustive calculations of the net profits for various setting BESS sizes. 
In addition, it has been found that updating the BESS size during the training process is beneficial to avoid failure in the training process of the bidding policy. 

Future work of this paper includes implementation and comparison of different variants of the proposed method based on continuous policy optimization algorithms for performance improvement. 
Another direction is to apply the proposed method to other co-optimization problems in power and energy fields.

\bibliographystyle{ieeetr}
\bibliography{ref}


\end{document}